\documentclass[fleqn,usenatbib]{mnras}
\usepackage{amsmath}	
\usepackage{amssymb}	

\usepackage{mathptmx}
\usepackage{txfonts}

\usepackage[T1]{fontenc}

\usepackage{graphicx}	

\newcommand \msun {M$_\odot$}

\newcommand \drvm  {$\Delta{\rm RV}_{\rm max}$}
\newcommand \fb {$f_{\rm bin}$}
\newcommand \kms {\,km\,s$^{-1}$}
\newcommand \fmIa {$f_{\textrm{merger} \rightarrow \textrm{Ia}}$}
\newcommand \fmwd {$f_\textrm{wd=merged}$}
\newcommand \dv {\mathrm{d}}

\defcitealias{Maoz_2012}{M12}
\defcitealias{Badenes_2012}{BM12}
\defcitealias{Maoz_2017}{MH17}

\title[Double-WD separation distribution and mergers]{The separation distribution and merger rate of double white dwarfs: improved constraints}

\author[D. Maoz, N. Hallakoun, C. Badenes]{
Dan Maoz$^{1}$\thanks{E-mail: \href{mailto:maoz@astro.tau.ac.il}{maoz@astro.tau.ac.il} (DM)},
Na'ama Hallakoun$^{1}$, and Carles Badenes$^{2,3}$
\\
$^{1}$School of Physics and Astronomy, Tel-Aviv University, Tel-Aviv 6997801, Israel\\
$^{2}$Department of Physics and Astronomy and Pittsburgh Particle Physics, Astrophysics and Cosmology Center (PITT PACC), University of Pittsburgh,\\ 3941 O'Hara Street, Pittsburgh, PA 15260, USA\\
$^{3}$Institut de Ci\`encies del Cosmos (ICCUB), Universitat de Barcelona (IEEC-UB), Mart\'i Franqu\'es 1, E08028 Barcelona, Spain
}

\date{Accepted XXX. Received YYY; in original form ZZZ}

\pubyear{2018}

\begin{document}
\label{firstpage}
\pagerange{\pageref{firstpage}--\pageref{lastpage}}\maketitle
\begin{abstract}
We obtain new and precise information on the double white dwarf (DWD) population and on its gravitational-wave-driven merger rate, by combining the constraints on the DWD population from two previous radial-velocity-variation studies: One based on a sample of white dwarfs (WDs) from the Sloan Digital Sky Survey (SDSS, which with its low spectral resolution probes systems at separations $a<0.05$\,au), and one based on the ESO-VLT Supernova-Ia Progenitor surveY (SPY, which, with high spectral resolution, is sensitive to $a<4$\,au). From a joint likelihood analysis, the DWD fraction among WDs is \fb$=0.095 \pm 0.020$ (1$\sigma$, random) $+ 0.010$~(systematic) in the separation range $\lesssim 4$\,au. The index of a power-law distribution of initial WD separations (at the start of solely gravitational-wave-driven binary evolution), $N\left(a\right)\dv a\propto a^\alpha \dv a$, is $\alpha=-1.30 \pm 0.15$ ($1\sigma$) $+0.05$~(systematic). The Galactic WD merger rate per WD is $R_\textrm{merge}=(9.7\pm 1.1)\times 10^{-12}$\,yr$^{-1}$. Integrated over the Galaxy lifetime, this implies that $8.5-11$~per cent of all WDs ever formed have merged with another WD. If most DWD mergers end as more-massive WDs, then some 10~per cent of WDs are DWD-merger products, consistent with the observed fraction of WDs in a ``high-mass bump'' in the WD mass function. The DWD merger rate is $4.5-7$ times the Milky Way's specific Type-Ia supernova (SN Ia) rate. If most SN~Ia explosions stem from the mergers of some DWDs (say, those with massive-enough binary components) then $\sim 15$~per cent of all WD mergers must lead to a SN~Ia.
\end{abstract}



\section{Introduction}
\label{sec:Intro}

The importance of mapping the demographics of the multiplicity of stars cannot be overstated. Such a mapping is required for understanding a myriad of astrophysical processes, from star formation, through binary evolution, binary interactions, supernovae, black hole formation, gravitational wave sources, and more. A promising approach toward this goal has been to try to characterise the binarity of stellar populations at different stages of their evolution --- whether when both stars are still on the main sequence \citep[e.g.,][]{Raghavan_2010}, when one member of the binary has evolved into a a giant \citep[e.g.,][]{Klein_2017,Badenes_2018} or a white dwarf (WD), \citep[e.g.,][]{RebassaMansergas_2017}, or when the binary system has evolved into a double WD (DWD) \citep[e.g.,][]{Maxted_1999, Maoz_2017}. DWDs are particularly interesting in numerous ways, including as possible progenitors of Type-Ia supernovae \citep[SNe Ia; e.g.][]{Maoz_2014} and as the main sources of low-frequency gravitational waves that will be detected by the next generation of space-based detectors \citep[e.g.,][]{Korol_2017}. DWDs are also attractive to study because their evolution is highly deterministic; the structure and the cooling processes of an individual WD in a DWD system are well understood and observationally tested \citep[e.g.,][]{Althaus_2010,Parsons_2017}, while the binary evolution, up to the final collision, is dictated by gravitational wave emission alone.

\citet[][hereafter M12]{Maoz_2012} developed a statistical method for characterizing observationally the binary population in a sample of stars, by means of \drvm, the maximum radial velocity (RV) shift observed between two or more epochs of the same star. The \drvm\ distribution can constrain the binary fraction of the population, \fb, within some separation, $a_\textrm{max}$, and the distribution of binary separations, $N\left(a\right) \dv a$. \citet[][hereafter BM12]{Badenes_2012} applied the \drvm method to a search for DWDs among $\sim 4000$ WDs from the Sloan Digital Sky Survey (SDSS). The coarse ($\sim 80$\kms) RV precision of the individual measurements (with a few epochs per WD) meant that these data probed the WD sample for DWDs at separations $a<a_\textrm{max}=0.05$\,au (effectively, only the 15 systems in the sample with observed \drvm$\gtrsim 250$\kms produced the ``signal'' in the \drvm\ distribution). \citetalias{Badenes_2012} found \fb\ to be in a rather broad range range of $3-20$~per cent ($1\sigma$). Assuming a power-law separation distribution, $N\left(a\right) \propto a^\alpha$, at the time of DWD formation (from which the binary separation evolves through gravitational-wave emission), \citetalias{Badenes_2012} constrained $\alpha$ to be from $-2$ to $+1$, but with a strong degeneracy between \fb\ and $\alpha$. Every combination of \fb\ and $\alpha$ gives a calculable specific WD merger rate for which, given the constraints above, \citetalias{Badenes_2012} obtained a merger rate per WD of $R_\textrm{merge}=2.6^{+6.3}_{-1.9}\times 10^{-12}$\,yr$^{-1}$.

More recently, \citet[][hereafter MH17]{Maoz_2017} applied the \drvm method to a sample of 439 WDs, each with few-epoch spectra from the European Southern Observatory (ESO) 8\,m Very Large Telescope (VLT), Supernova-Ia Progenitor surveY \citep[SPY;][]{Napiwotzki_2001}. Although the sample size was an order of magnitude smaller than the SDSS sample of \citetalias{Badenes_2012}, therefore lowering SPY's sensitivity to rare, small-separation DWDs with high RVs, the typical RV resolution of these data, $1-2$\kms\ per epoch ($\sim 50$ times better than SDSS), combined with the distribution of time separations between epochs, meant that the SPY sample is sensitive to DWDs out to separations $a_\textrm{max}=4$\,au. The SPY sample effectively probes for DWDs in the $a=0.05-4$\,au range, complementing the \citetalias{Badenes_2012} SDSS study that constrained binarity in the $a=0.001-0.05$\,au range (the lower limit in SDSS arising from the individual exposure length of $\sim 15$\,min). From the SPY analysis, \citetalias{Maoz_2017} found a fraction of WDs that are DWDs, with separations $a<4$\,au, in the range \fb$=0.100 \pm 0.020$ (1$\sigma$, random) $+ 0.02$~(systematic, arising mainly from uncertainty regarding the WD mass function). The power-law index of the DWD separation distribution at birth was $\alpha=-1.3 \pm 0.2$ ($1\sigma$) $\pm 0.2$~(systematic), and the specific merger rate of DWDs is $R_\textrm{merge}=(1.1 \pm 0.3) \times 10^{-11}$\,yr$^{-1}$ per WD, but with a broad allowed $2\sigma$ range, from $3 \times 10^{-12}$\,yr$^{-1}$ to $3 \times 10^{-10}$\,yr$^{-1}$ per WD. Follow-up efforts to characterize the individual candidate DWDs identified by \citetalias{Badenes_2012} and \citetalias{Maoz_2017} include \citet{Breedt_2017}.

In this Letter, we combine the SDSS and SPY results in a joint likelihood analysis, to obtain the best constraints to date on the DWD population and its merger rate. The new analysis implies that some $10$~per cent of the WDs in the Milky Way (MW) have undergone a merger with another WD over the lifetime of the Galaxy, possibly explaining the ``bump'' at $\sim 1$\,\msun, observed in the WD mass function by some studies. The WD merger rate that we find is about $6$ times the SN~Ia rate in the Galaxy. This indicates that, if most SNe~Ia arise from DWD mergers, then some $15$~per cent of such mergers lead to a SN~Ia explosion.

\section{Joint SDSS-SPY constraints on binary population parameters}
\label{sec:Joint}

We briefly review the essentials of the \drvm\ method and its use to constrain the DWD population. See \citetalias{Maoz_2012} for a detailed exposition, or \citetalias{Maoz_2017} for an intermediate-length version. For a sample of objects, say WDs, having multiple-epoch RV measurements, the observed \drvm\ statistic is the distribution of maximum RV differences between any two epochs for every object. To find the range of model WD populations that are consistent with the observed \drvm\ distribution, a grid of simulated model populations is created, the grid spanning the parameters that describe the population. A particular WD population model consists of a large number [$O(10^5)$] of WDs. Each simulated WD has a probability \fb\ to be in a DWD system with separation within $a_\textrm{max}$, and probability $1-$\fb\ to be single. If an initial draw determines the simulated WD is in a binary, the physical DWD system parameters are then chosen: the mass $m_1$ of the WD (drawn from an observationally informed mass function); the mass $m_2$ of the companion WD (drawn from a mass-ratio distribution); and the binary separation $a$, drawn from a separation distribution $N\left(a\right) \dv a$. It is assumed that, at the time they first become close DWDs, after their final common-envelope phase, the DWD distribution of separations can be approximated as a power law with index $\alpha$. This parameter is physically set by combined effects of the zero-age-main-sequence binary separation distribution and by common-envelope evolution. We assume $\alpha$ is independent of time (along the Galaxy's history) and of WD mass. This assumption is not necessarily valid, but our observational measurement of $\alpha$, supposing the assumption is valid, can be used as a ground truth to be matched by binary population synthesis models and models of common-envelope evolution. Orbital decay due to gravitational wave emission will modify this initial separation distribution with time, and furthermore, the DWD separation distribution today will be the combination, over the lifetime of the Galaxy, of the evolved distributions from many populations of different ages. \citetalias{Maoz_2012} showed that this results in a separation distribution
\begin{equation}
N(x)\propto x^{4+\alpha} [(1+x^{-4})^{(\alpha+1)/4}-1],~~~~~ \alpha\ne -1,
\label{eqtimeint}
\end{equation}
or
\begin{equation}
N(x)\propto x^{3} \ln(1+x^{-4}),~~~~~ \alpha= -1,
\label{eqtimeintlog}
\end{equation}
where
\begin{equation}
x\equiv\frac{a}{(K t_0)^{1/4}},
\end{equation}
\begin{equation}
\label{eq:K}
K\equiv\frac{256}{5}\frac{G^3}{c^5}m_1 m_2 (m_1+m_2),  
\end{equation}
$t_0$ is the age of the Galaxy, and $G$ and $c$ have their usual meanings. $x=1$ (i.e. ${a}={(K t_0)^{1/4}}$), corresponds to the separation of DWDs that merge within the lifetime of the Galaxy. The present-day distribution $N\left(x\right) dx$ is approximately a broken power law, with index $\alpha$ at $x \gg 1$. At $x \ll 1$ the power-law index is 3 for $\alpha\ge -1$, and $\alpha + 4$ for $\alpha\le -1$. For each simulated DWD system, $a$ is drawn from $N(x)$, between $a_\textrm{min}=2\times 10^4$\,km (DWD contact) and $a_\textrm{max}$. Kepler's law then gives the period and the (assumed circular) orbital velocities. The merger rate per WD, for a given set of WD population parameters, is found numerically as the fraction of all of the systems in the simulation whose merger lifetime, $t_\textrm{merge}=a^4/K$ is within some time interval, divided by that time interval. The DWD merger rate per WD, divided  by the MW ratio between stellar mass and WD number ($15.5\pm 1.8$\,\msun\ per WD, obtained from the local WD number density, $0.0055\pm 0.0001$\,pc$^{-3}$, \citealt{Munn_2017}, and the stellar mass density, $0.085\pm 0.010$\,\msun\,pc$^{-3}$, \citealt{McMillan_2011}, see \citetalias{Maoz_2017}) gives, for a particular simulated WD population model, the DWD merger rate per unit stellar mass.

Every simulated DWD system is then ``observed'', by drawing for it a line-of-sight inclination of the orbital plane, choosing which of the two WDs is the photometric primary whose RV will be measured, and sampling the line-of-sight velocity (i.e. RV) curve of the photometric primary with a particular observation pattern (number of epochs and time between epochs) chosen at random from the set of real observing patterns of the real observed sample. Every simulated velocity measurement is noised with a random RV error, drawn from the distribution of measurement errors of the observed sample. The $1-$\fb\ fraction of simulated WDs, that are effectively single, skip directly to this stage of the simulation, where they are assigned a chosen number of epochs, and at each epoch a random RV error around zero velocity. To obtain the model's \drvm, we find the difference between the minimum and maximum RV for every simulated DWD or single WD. The fractional prediction for each bin in the model \drvm\ distribution, multiplied by the observed WD sample size, gives the expectation value for that bin. To compare each simulated model \drvm\ distribution to the observed distribution, we take the model expectation value for each velocity bin in the \drvm\ distribution, and we sum the logarithms of the Poisson probabilities of finding the observed number of systems in each bin, given the expectations from the model. This gives the log of the likelihood of each model. By running models over a grid in parameter space, we find the allowed region of the DWD population parameter space.

The main result of \citetalias{Badenes_2012}, based on analysis of the SDSS WD sample, was to constrain the DWD population parameters to a region of parameter space consisting of the power-law index of the separation distribution at birth, $\alpha$, and $f_\textrm{bin, 0.05}$, the fraction of WDs that are in binaries with separations $a < 0.05$\,au. \citetalias{Maoz_2017}, in turn using the SPY sample, found the allowed region in the space of $\alpha$ and of  $f_\textrm{bin, 4}$, the fraction of WDs in binaries with separations $a < 4$\,au. Before combining the results of these two studies in a joint likelihood analysis, we need to bring them to a common parameter plane by, e.g., mapping the two-dimensional likelihood function
\begin{equation}
L(f_\textrm{bin, 0.05}, \alpha)\rightarrow  L(f_\textrm{bin, 4}, \alpha)
\end{equation}
by means of the transformation 
\begin{equation}
f_\textrm{bin, 4}=\frac{\int_{a_\textrm{min}}^{4\,\textrm{au}} N(a, \alpha) \dv a}{\int_{a_\textrm{min}}^{0.05\,\textrm{au}} N(a, \alpha) \dv a}f_\textrm{bin, 0.05}.
\end{equation}
For this calculation we used $N(x)$ from Equations \ref{eqtimeint}$-$\ref{eq:K}, with a MW lifetime of $t_0=10$\,Gyr, and with the mean values of $m_1$ and $m_2$ from the simulations, which gives a turnover separation in the distribution, i.e. $x=1$, that corresponds to $a=0.0107$\,au. Since the transformation differs for different values of $\alpha$, it distorts the likelihood contours, stretching them more toward higher \fb\ the higher the value of $\alpha$. Finally, once the likelihoods from the two studies are in the same parameter space, we obtain the joint likelihood by simply summing the two log-likelihoods at each point in parameter space. 

\section{Results}
\label{sec:Results}

\begin{figure}
\centering
\includegraphics[width=\columnwidth]{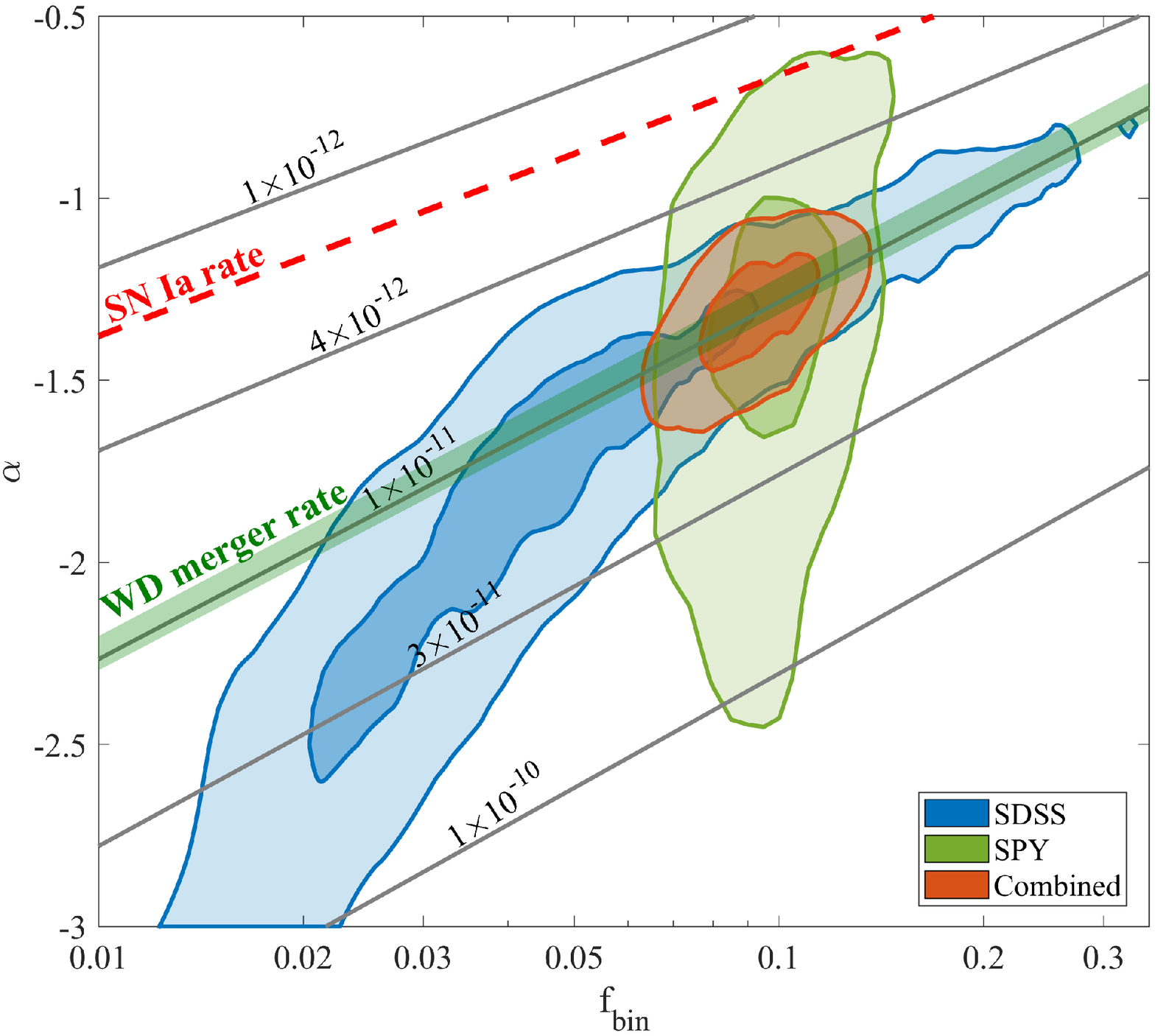}
\vspace{0.02\textheight}
\includegraphics[width=\columnwidth]{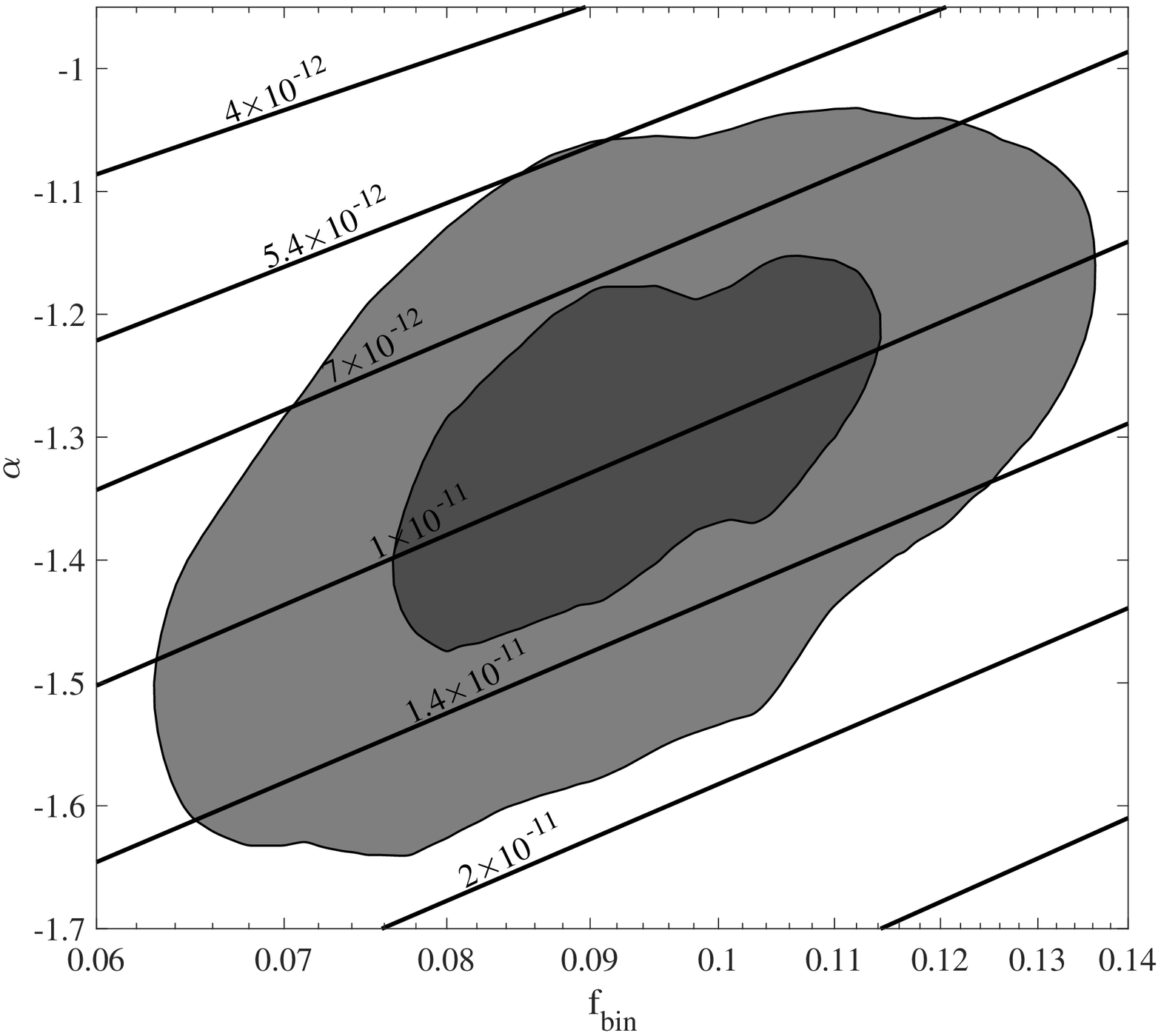}
\caption{\textit{Top:}  $1\sigma$ and $2\sigma$ likelihood contours in the plane of $f_\textrm{bin, 4}$, the fraction of WDs in binaries with separations less than 4~au, and  $\alpha$, the power law index of the initial  DWD separation distribution, for the SDSS WD sample from \citetalias{Badenes_2012}, after transforming their results from their ($f_\textrm{bin, 0.05}, \alpha$) parameter space (blue); likelihood contours for the SPY sample from \citetalias{Maoz_2017} (green); and joint likelihood contours from combining the two sets of results (red). Straight lines are loci of constant DWD merger rate, as marked in units of mergers per year per WD. Also marked are a line showing the SN-Ia rate in a MW-like galaxy (red), and a band with the best fit likelihood weighted 1$\sigma$ range for the WD merger rate (green). \textit{Bottom:} zoom in on the joint likelihood distribution alone. Integrated over a MW lifetime of $10^{10}$\,yr, the allowed values of the merger rate imply that about $10$~per cent of WDs have undergone a DWD merger. The SN Ia rate per WD in the MW is $R_\textrm{Ia}\approx 1.7\times 10^{-12}$\,yr$^{-1}$, one-sixth the DWD merger rate. If most SN Ia explosions involve a DWD merger, then about one in seven mergers lead to a SN Ia.}
\label{fig:Contours}
\end{figure}

Figure~\ref{fig:Contours} shows, for the transformed results of \citetalias{Badenes_2012}, for the results of \citetalias{Maoz_2017}, and for the new joint analysis presented here, the parameter space of $f_\textrm{bin, 4}$ and $\alpha$, with contours at the $1\sigma$ and $2\sigma$ likelihood levels, corresponding to distances of $0.5$ and $2$ in log-likelihood from the best-fit model. (We note that \citetalias{Badenes_2012} drew these respective contours at $1$ and $4$, and we correct this oversight here.) Remarkably, the different orientations of the likelihood contours from the two studies lead, in the joint analysis, to a rather small allowed region in parameter space. The $1\sigma$ range for \fb\ is \fb$=0.095 \pm 0.020$, ($+0.01$, considering the systematic error in \citetalias{Maoz_2017}), not too different from what was found by \citetalias{Maoz_2017}. This similarity arises from the fact that the \citetalias{Maoz_2017} contours are fairly vertically oriented in this plane. Their narrowness arises from the high RV resolution of SPY, which efficiently detects most of the $\sim 1$\,au-separation DWDs. The contour elongation in the $\alpha$ direction results from the smallness ($\sim 400$~WDs) of the SPY sample, which prevents the discovery of the rare small-separation DWDs; their numbers are therefore poorly constrained by SPY, and hence so is $\alpha$.  

However, the intersection of these contours with the diagonally oriented contours from \citetalias{Badenes_2012} now strongly constrain $\alpha=-1.30 \pm 0.15$ ($+0.05$ systematic). The straight lines seen in Figure~\ref{fig:Contours} are loci of constant merger rate (see \citetalias{Maoz_2012} for an explanation of their approximately linear form in a plot such as this one, with axes that are linear in $\alpha$ and logarithmic in \fb). The tight joint constraints on \fb\ and $\alpha$ lead to a $1\sigma$ range for the likelihood-weighted merger rate of the DWD population of $R_\textrm{merge, WD}=(9.7\pm1.1)\times 10^{-12}$\,yr$^{-1}$ per WD. To obtain the DWD merger rate per unit stellar mass, we divide by $(15.5\pm 1.8)$\,\msun\ per WD in the Galactic disk (see above), and obtain $R_\textrm{merge, m*}=(6.3\pm 1.0)\times 10^{-13}$\,yr$^{-1}$\,\msun$^{-1}$.

\citet{Toonen_2017} have recently studied WD binarity in a local 20\,pc sample, and compared it to their binary population synthesis predictions. Among their findings, they estimate that the observed fraction of WDs in angularly ``resolved DWD'' systems, which in their sample corresponds to separations of roughly $10^1-10^4$\,au, is about $1.5-2$~per cent, an order of magnitude less than \citet{Toonen_2017} expect from their models. If the power-law separation distribution that we have assumed extends out to large separations with the same best-fit index that we have found, $\alpha\approx -1.3$, then the binary fraction at $<4$\,au that we have measured, $8.5-11$~per cent, corresponds to a factor of $\approx 7$ fewer DWD systems at $10^1-10^4$\,au, i.e. $1-1.6$~per cent, in good agreement with the observed resolved DWD fraction. In other words, irrespective of any conflicts with model binary population model expectations, the observed DWD fractions at $<4$\,au and at large separations are consistent with a single distribution having the power-law slope and normalization that we have found.

\section{Implications}
\subsection{The DWD merger history and mass function}
As a rule, observed WD mass functions are sharply peaked, with the large majority of WDs concentrated at masses around $\approx 0.6\pm 0.2$\,\msun\ \citep[e.g.][]{Giammichele_2012, Tremblay_2016}. In a merger of two typical-mass WDs at the conclusion of a slow in-spiral driven by gravitational-wave emission, the less-massive (and hence less dense) WD will likely be tidally sheared into an accretion disc around the more-massive WD. Following any accretion-related processes, or after some episode of revived stellar evolution caused by the addition of mass and nuclear fuel to the more-massive WD, the merged system will settle down as a WD having the combined mass of the DWD system, minus any mass loss that occurred at any stage \citep{Shen_2012, Shen_2013, Dan_2014}. The present-day DWD merger rate per WD in the solar neighbourhood, found above, is $R_\textrm{merge, WD}=(9.7\pm1.1)\times 10^{-12}$\,yr$^{-1}$. Assuming a roughly constant star-formation rate over most of the MW lifetime of $\sim 10^{10}$\,yr, the DWD merger rate will also be constant (see \citetalias{Maoz_2012}). This implies that, integrated over the MW lifetime, some $10\pm 1$~per cent of WDs in the Galaxy have merged with another WD. If, indeed, most DWD mergers end up as massive merged WDs, the WD mass function might display some prominent feature revealing this merged, high-mass, WD population.

Past reports of such a ``high-mass bump'' in the WD mass function have come already from \citet{Marsh_1997} and \citet{Vennes_1999}. More recently, \citet{Liebert_2005} took the WD mass distribution obtained from their flux-limited Palomar-Green sample of WDs, and corrected it to account for the survey's lower sensitivity to massive WDs (which have smaller surface areas and are therefore detected out to a smaller volume than typical-mass WDs). They found (see their fig. 13) a prominent component in the mass function at masses of $\approx 0.75-0.95$\,\msun, constituting some 20~per cent of the population. \citet{Koester_2014} show in their fig. 1 the mass function for the small sample of WDs that they chose for \textit{Hubble Space Telescope} ultraviolet spectroscopy in search of atmospheric metals. Although this is an unbiased selection of WDs from, again, a flux-limited sample (with an additional criterion on effective temperature, $17\,000$ to $27\,000$\,K), even without a survey-volume correction the mass function shows a bump at $\approx 0.8$\,\msun, with $14/85$ ($16$~per cent) of the WDs in this component. As opposed to these flux-limited samples, \citet{Giammichele_2012} obtained the mass function for a volume-limited sample of $169$ WDs within $20$\,pc. The mass function has $23$ WDs (i.e. $14$~per cent of the sample) above a mass of $0.8$\,\msun, with $6$~per cent at about $1$\,\msun, a component that \citet{Giammichele_2012} conclude must consist of DWD-merger products.

In contrast, \citet{Kepler_2007, Hu_2007} and \citet{DeGennaro_2008} analysed samples of WDs from SDSS having $T_\textrm{eff} \gtrsim 12\,000-13\,000$\,K, and found mass functions that show little or no evidence for a high-mass bump. \citet{RebassaMansergas_2015}, however, obtained the mass function from an updated sample of SDSS WDs \citep{Kepler_2015}, but including also faint WDs (absolute bolometric magnitudes of $12-13$) with $6\,000<T_\textrm{eff} < 12\,000$\,K WDs. \citet{RebassaMansergas_2015} argue that, as opposed to previous analyses that excluded the cool SDSS WDs, their analysis can include them, because the atmospheres of these objects can now be properly modelled by applying 3D model corrections (\citealt{Tremblay_2013}, previously unavailable) to the 1D model treatment of convective energy transport. As a result, the corrected models give reliable estimates of surface gravity, and hence of mass, for the faint and cool WDs. Inclusion of these WDs in the sample leads to a prominent high-mass bump in the \citet{RebassaMansergas_2015} WD mass function, with about $20$~per cent of WDs at $0.8-1.0$\,\msun. A caveat previously raised by \citet{Kepler_2013}, however, is that apart from the 3D treatment of convection, the masses of cool WDs might still be over-estimated if unresolved Zeeman splitting by weak magnetic fields increases the apparent widths of the Balmer lines, mimicking the stronger Stark broadening of a higher surface gravity. Finally, \citet{Tremblay_2016} have produced the latest SDSS-based WD mass function. They discuss at length whether there is a high-mass merger-related excess in their mass function. They conclude that there in none, and even find a deficit of massive WDs compared to expectations from simulations. However, their sample excludes cool ($<16\,000$\,K) WDs. If \citet{RebassaMansergas_2015} are correct that the high-mass bump consists mainly of cool WDs, then the deficit is not surprising.

\begin{figure}
\centering
\includegraphics[width=\columnwidth]{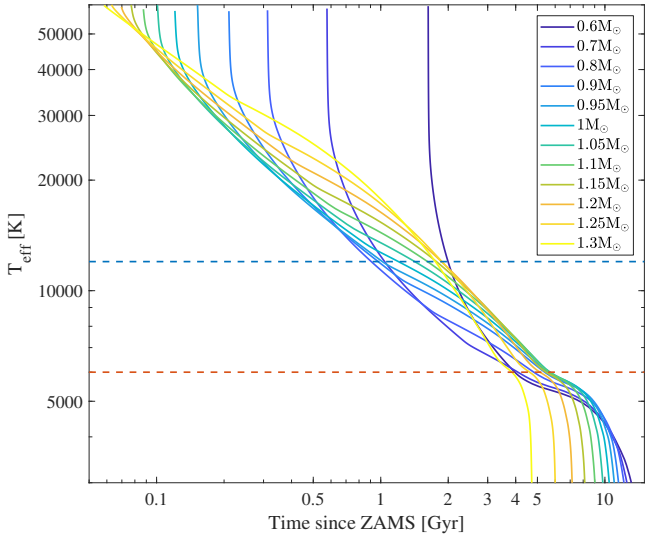}
\caption{WD cooling curves for thick hydrogen envelope from \citet{Fontaine_2001}, as a function of the total age since the birth of the WD progenitor. Different colours account for different WD masses. The blue (red) dashed line marks an effective temperature of $12\,000$\,K ($6000$\,K).}
\label{fig:Cooling}
\end{figure}

To further examine the question of how temperature cuts in a sample may exclude massive WDs, we plot in Fig.~\ref{fig:Cooling} the theoretical cooling curves of WDs of different masses from \citet{Fontaine_2001}~\footnote{\url{http://www.astro.umontreal.ca/~bergeron/CoolingModels}}, showing effective temperature as a function of the total age since the zero-age main sequence (ZAMS) of the WD progenitor stars. The time since ZAMS until the formation of the WD is calculated for solar metallicity using the single star evolution (\textsc{sse}) code of \citet{Hurley_2000,Hurley_2013}. The progenitor mass is related to the WD mass through the initial-final mass relation of \citet{Tremblay_2016}. We choose cooling curves for carbon-oxygen WD models with thick surface hydrogen envelopes, given that the mass-radius relation of WDs in detached eclipsing  binaries, as observed  by \citet{Parsons_2017}, is in excellent agreement with such models. Fig.~\ref{fig:Cooling} illustrates how, although more massive WDs cool more slowly (shallower slopes in the top-left part of the figure, a result of their lower surface areas), their shorter progenitor lifetimes mean that the WD temperatures fall below $12\,000$\,K, sooner after star formation than those of less massive WDs. In particular, compared to the common $0.6$\,\msun\ WDs, WDs of mass $0.7-1.0$\,\msun\ take about half the time since star formation to cool below this temperature. In a $T>12\,000$\,K WD sample from an environment that has had a constant star formation rate over the past few Gyr, massive WDs will be under-represented by a factor of about 2. Conversely, over 10\,Gyr after star formation, most WDs remain warmer than $5000$\,K, and it is only the very massive, $> 1$\,\msun\ WDs that undergo crystallisation and subsequent accelerated Debye cooling to lower temperatures \citep[see][]{Fontaine_2001}. Thus, samples that include WDs down to $\sim 5000$\,K should contain most of the mass range of single WDs produced over the age of the Galaxy, with only the most massive, $> 1$\,\msun, WDs under-represented to some extent. That said, we note that this picture is an over-simplification over what is needed in the present context, since WDs can be reheated by binary evolution, including common envelope phases, and certainly by DWD mergers. It is therefore unclear whether the high-mass excess among cool SDSS WDs, shown by \citet{RebassaMansergas_2015}, is real or not. 

Synthesizing, it appears that, in many samples, the WD mass function does have a high-mass component that includes roughly $15-20$~per cent of all WDs. If the high-mass bump is not the result of over-estimated masses for some WDs (e.g. due to weak magnetic fields producing line broadening through unresolved Zeeman splitting), then this observed fraction is in rough agreement with the fraction 
expected from the DWD merger rate that we have estimated here.         

\subsection{The Type-Ia supernova progenitor population}
For the past several decades, two main categories of progenitor models for SNe~Ia have been the so-called single-degenerate and double-degenerate scenarios (see \citealt{Maoz_2014} for a review). Among the various problems they face, each scenario is challenged in terms of its predicted numbers of progenitor systems, compared to the known or observed numbers of such systems. In single-degenerate models, where the  progenitors are WDs growing in a mass through stable Roche-lobe accretion from normal-star binary companions, one needs to accrete of order a solar mass of material on to the WD (to bring it from typical WD masses to near the Chandrasekhar mass), at an accretion rate of $\sim 10^{-7}$\,\msun\,yr$^{-1}$ \citep[to achieve stable hydrogen burning, and accumulation of helium ash on the WD surface;][]{Nomoto_1982, Shen_2007}. The accretion phase must therefore last $\sim 10^{7}$\,yr. The MW hosts about one SN~Ia per $200$\,yr (e.g. \citealt{Maoz_2014}). Our Galaxy should therefore currently host $\sim 10^5$ accreting WDs that are growing toward the Chandrasekhar mass and toward their explosions as SNe~Ia. This is in stark contrast to the known numbers, of order only $10$, of systems thought to be candidates for such accreting WDs, namely recurrent novae and super-soft X-ray sources. A similar discrepancy of several orders of magnitude results from such an accounting in other Local Group galaxies (see \citealt{Maoz_2014}.)      
   
In the double-degenerate picture, the SN~Ia progenitor systems are close DWDs that merge. Naturally, the Galaxy's DWD merger rate must be at least as large as its SN~Ia rate, if mergers are to produce most SNe~Ia through this channel. However, as already mentioned above, it has long been recognized that most DWD mergers will produce not a SN Ia explosion, but more likely a more-massive, merged, WD. Even if a thermonuclear detonation is somehow always ignited in the core of one of the WDs undergoing a merger, combustion of about $0.5$\,\msun\ of WD material into $^{56}$Ni, as indicated by basic SN~Ia energetics and observations, requires a WD core density of $\gtrsim 2 \times 10^7$\,g\,cm$^{-3}$, which exists only in single WDs of $\gtrsim 0.9$\,\msun\ \citep[B. Katz, private communication;][]{Moll_2014}. Since massive WDs are the exception, rather than the rule, among the WD population, this means that, in the double-degenerate scenario, only a fraction of DWD mergers lead to a SN~Ia, i.e. those in which at least one of the WDs is massive enough to have the critical core density. This, in turn, means that for the scenario to be viable, the DWD merger rate must be significantly larger than the SN~Ia rate. However, the Galaxy's DWD merger rate cannot be arbitrarily high, since given the observed number of WDs and the age of the Galaxy, all WDs would have then merged by now (and would have  become either bigger WDs or SNe~Ia). These coupled constraints on the fraction of DWD mergers that result in a SN~Ia explosion, \fmIa, and on the fraction of the present WD population that are DWD merger products, \fmwd, can be expressed as
\begin{equation}
\label{eq:Mergers}
f_\textrm{wd=merged} 
\frac{f_{merger\rightarrow Ia}}{1-f_{merger\rightarrow Ia}} = R_\textrm{Ia, wd}\times t_0, 
\end{equation}
where $R_\textrm{Ia, wd}$ is the MW's SN~Ia rate per WD, and $t_0$ is the Galaxy's age. Taking the SN~Ia rate per unit stellar mass for an Sbc galaxy of MW mass from \citet{Li_2011}, $R_\textrm{Ia, m}= 1.1 \times 10^{-13}$\,yr$^{-1}$\,\msun$^{-1}$, and multiplying by a stellar mass to WD ratio of $(15.5\pm1.8)$\,\msun\ per WD (see above) and by $t_0\approx 10^{10}$\,yr gives $R_\textrm{Ia, wd}\times t_0=0.017\pm 0.02$. In other words, from the measured SN~Ia rate in MW-like galaxies and the estimated ratio of stellar mass to WDs in the MW, $1.5$ to $2$~per cent of the WDs that formed in our Galaxy have already exploded as SNe~Ia. If most SNe~Ia explode through the double-degenerate channel, then Equation~\ref{eq:Mergers} constrains their progenitor systems: the product of the fraction of WD mergers that lead to a SN~Ia and the fraction of present day WDs that are merger remnants must equal the $1.5-2$~per cent fraction of all WDs that have exploded as SNe~Ia.
 
The DWD merger rate that we have found, $(9.7\pm1.1)\times 10^{-12}$\,yr$^{-1}$, is $5.7\pm 1.2$ times the Galaxy's SN~Ia rate per WD, $R_\textrm{Ia}\approx (1.7\pm 0.2)\times 10^{-12}$\,yr$^{-1}$, and therefore, if all SNe~Ia come from DWD mergers, then \fmIa$=0.15\pm 0.03$. Thus, the product constraint in Eq.~\ref{eq:Mergers} is satisfied by the combination of this fraction of mergers that lead to a SN~Ia explosion, and a $f_\textrm{wd=merged} = 0.097\pm 0.011$ fraction of WDs that have merged. As noted above, this merged fraction seems roughly consistent with the fraction in the high-mass tail of the WD mass function. However, the $\sim 12-18$~per cent of mergers that end in a SN~Ia may be uncomfortably high. At first, one might think that this a natural outcome of the fact that some $15-20$~per cent of the WD mass function is in massive WDs, and therefore among all DWDs, a similar fraction contains a massive member, massive enough to attain the WD core density needed for a successful explosion. One must remember however, that at least some, perhaps all, of these massive WDs appear to be the already-merged products of lower-mass WDs. In order to utilize the massive WDs in a subsequent merger with a third WD, that happens over the Galaxy lifetime to, say, $10$~per cent of the systems (just to mimick what we have found for DWDs), then to have $1.5-2$~per cent of all WDs explode as SNe~Ia, means that most of the merged DWDs need to have originated in triple-star systems. However, among main-sequence A-type stars that are the progenitors of typical WDs, although the binary fraction is high and approaches unity, the triple fraction is only $10-20$~per cent \citep{Duchene_2013, Leigh_2013}. This suggests a remaining factor-of-few discrepancy between the rate of SNe~Ia and the estimated number of progenitor systems expected in the double-degenerate scenario. The tension could be relieved if either the MW's SN~Ia rate is over-estimated, or the WD number density is underestimated, both of which would lower the right-hand side of Eq.~\ref{eq:Mergers}. A smaller fraction \fmIa\ of mergers would then need to explode as SNe~Ia, a fraction perhaps attainable by means of triple systems in which the inner close binary has merged into a massive WD, and is then detonated as a SN~Ia when it eventually merges with the tertiary WD. In any event, it is worth remembering that if a merger of a DWD with a WD more massive than $0.9$\,\msun\ is assumed to lead to a SN Ia, then the double-degenerate progenitor population problem is at the level of a factor of a few, while for single degenerates it is at orders of magnitude.

Progress on these questions will come soon from the census of WDs being carried out by the \textit{Gaia} Mission, including single WDs, DWDs, and WDs in binaries with main-sequence stars. The picture will further clear up with the massive spectroscopic followup of WDs, identified by \textit{Gaia}, by the upcoming ($2020$) fifth incarnation of SDSS \citep{Kollmeier_2017}.

\section*{Acknowledgements}
We acknowledge the hospitality of the Peking Kavli Institute of Astronomy and Astrophysics, and hosts Subo Dong and Zhuo Li of the workshop on Transients from Compact Objects in October 2017, where this work was suggested by the participants. Boris G\"{a}nsicke and Boaz Katz are thanked for discussions and advice.
This work was supported by Grant 1829/12 of the Israeli Centers for Research Excellence (I-CORE) programme of the Planning and Budgeting Committee (PBC) and the Israel Science Foundation (ISF). This research has made use of the VizieR catalogue access tool, CDS, Strasbourg, France. The original description of the VizieR service was published in A\&AS 143, 23.




\bibliographystyle{mnras}
\bibliography{wdjoint} 




\appendix

\section{DWD candidates from the SDSS}
To complement the list of candidate DWDs from SPY, published in table~1 of \citetalias{Maoz_2017}, we list here the candidate DWD systems from the tail of the \drvm\ distribution of SDSS WDs in \citetalias{Badenes_2012}, and describe any additional information that is currently known about these systems.

\begin{table}
\caption{Candidate DWDs with \drvm > 250\kms.}
\label{tab:Tail}
\begin{center}
\begin{tabular}{l r r c}
\hline
\multicolumn{1}{l}{Name} &
\multicolumn{1}{c}{\drvm} & 
\multicolumn{1}{c}{$M_1$} &
\multicolumn{1}{l}{Comments} \\
&
\multicolumn{1}{c}{[km\,s$^{-1}$]} &
\multicolumn{1}{c}{[\msun]} & \\
\hline
SDSS\,J125733.64+542850.5 & 537.95 & 0.15 & 1\\
SDSS\,J092345.60+302805.0 & 412.11 & 0.23 & 2\\
SDSS\,J041518.91+165238.2 & 380.26 & 0.50 & \\
SDSS\,J234902.80+355301.0 & 378.43 & 0.39 & 3\\
SDSS\,J034319.09+101238.0 & 336.63 & 0.89 & \\
SDSS\,J011100.64+001807.1 & 335.06 & 0.79 & 4\\
SDSS\,J011302.70$-$100615.9 & 323.64 & 0.62 & \\
SDSS\,J104524.61+414439.8 & 313.03 & 0.40 & \\
SDSS\,J165923.87+643809.3 & 290.63 & 0.75 & \\
SDSS\,J210308.80$-$002748.8 & 290.17 & 0.17 & 5\\
SDSS\,J033205.53+011206.7 & 287.36 & 0.64 & \\
SDSS\,J100554.06+355014.2 & 284.24 & 0.17 & 6\\
SDSS\,J151338.83+573920.1 & 283.56 & 0.65 & \\
SDSS\,J111501.16$-$124217.9 & 281.58 & 0.51 & \\
SDSS\,J012534.99+385046.9 & 255.68 & 0.58 & \\
\hline
\end{tabular}
\end{center}

\begin{flushleft}
Notes:
$M_1$ is the derived mass for the photometric-primary WD from \citet{Kleinman_2013}, except for the systems with a full orbital solution, where $M_1$ was taken from the references mentioned below.\\
\vspace{0.1cm}
(1)~SDSS\,J125733.64+542850.5: $P=4.56$\,h, $M_1=0.15$\,\msun, $M_2=1.06$\,\msun\ \citep{Badenes_2009, Kulkarni_2010, Marsh_2011b, Bours_2015}\\
(2)~SDSS\,J092345.60+302805.0: $P=1.08$\,h, $M_1=0.23$\,\msun, $M_2\leq0.34$\,\msun\ \citep{Brown_2010}\\
(3)~SDSS\,J234902.80+355301.0: $P=4.35$\,h, $M_1=0.39$\,\msun, $M_2\leq0.74$\,\msun\ \citep{Breedt_2017}\\
(4)~SDSS\,J011100.64+001807.1: pulsating ZZ Ceti, has \textit{K2} data \citep[EPIC\,229228480;][]{Hermes_2017}\\
(5)~SDSS\,J210308.80$-$002748.8: $P=4.87$\,h, $M_1=0.17$\,\msun, $M_2\leq0.71$\,\msun\ \citep{Kilic_2012}\\
(6)~SDSS\,J100554.06+355014.2: $P=4.24$\,h, $M_1=0.17$\,\msun, $M_2\leq0.19$\,\msun\ \citep{Kilic_2012}

\end{flushleft}
\end{table}

%
%


\bsp	
\label{lastpage}
\end{document}